# A novel process to produce amorphous nano-sized B useful for $MgB_2$ phase synthesis.

**Maurizio Vignolo[1*], Gennaro Romano[1], Cristina Bernini[1], Alberto Martinelli[1] and Antonio Sergio Siri[1,2]**

1 CNR-SPIN, C.so Perrone 24 canc., 16152 Genoa - Italy
2 DiFi, Physics department of the University of Genoa, Via Dodecaneso 33, 16132 Genoa - Italy

*Corresponding author: tel. +39-010- 6598790; e-mail: maurizio.vignolo@spin.cnr.it

Abstract: In this paper we report a new synthesis route to produce boron powders characterized as being amorphous and having very fine particle size. This route has been developed to improve the performances of superconducting $MgB_2$ powders, which can be directly synthesized from this nano-structured boron precursor by following the *ex-situ* or the *in-situ* P.I.T. method during the manufacturing of tapes, wires and cables. All the procedure steps are explained and the chemical-physical characterization of the boron powder, using x-ray diffraction (Xrd), scanning electron microscopy (SEM) and transmission electron microscopy (TEM) techniques, is reported. Furthermore, a comparison with commercial boron is given. Preliminary results of the magnetic and electrical characterization, such as the critical temperature ($T_C$) and the transport critical current density ($J_{C\,t}$), for the $MgB_2$ tape are reported and compared with the tape prepared with commercial boron.

I. Introduction

Since the discovery of the superconducting properties in magnesium diboride ($MgB_2$) all the studies and the investigations on this material were focused on using either the pre-reacted commercial $MgB_2$ or using commercial precursors, boron (B) and magnesium (Mg). Many efforts to improve the performances of wires [1], tapes [2] and cables [3] have been done; among the most pursued routes there is the critical current density ($J_c$) improvement by doping $MgB_2$ powder with various elements [4], [5] and compounds [6], [7]. However, it is always a challenge to reach a homogeneous dispersion of the dopant both at the crystallographic level, at the boron site, as well as dispersed into $MgB_2$. In fact, the elemental B as well as most of the dopants are not suitable to be manipulated: i.e. boron, carbon and SiC are characterized by high hardness and melting point.

There are two main techniques to obtain pure boron powder: via-gas [8], [9] and magnesiothermic reaction [10]. The first technique uses the hydrogen gas to reduce gaseous boron trichloride ($BCl_3$), whereas the latter uses a combination of heat and magnesium to reduce the boron trioxide ($B_2O_3$) to elemental boron. The gaseous route is the most dangerous, involving the reaction between two highly reactive gases ($H_2$ and $BCl_3$) and the formation of a highly corrosive gas (HCl) according to the reaction: $2BCl_{3(g)}+3H_{2(g)}=6HCl_{(g)}+2B_{(s)}$. The conditions are even more complicated in terms of safety and health when B is obtained by thermal decomposition of diborane ($B_2H_6$): $B_2H_{6(g)}=2B_{(s)}+3H_{2(g)}$. In fact, this gaseous boron precursor must be handled with care since is highly reactive, and catches flame spontaneously in air or explode with pure oxygen; therefore it is also used as rocket propellant. Furthermore, it is very dangerous for human health because of its poisonous and toxic effects on the central nervous system. In standard condition the gas route gives very pure crystalline boron (≥99%), whereas the metallothermic route produces boron with lower purity (90-97%). The main impurities affecting the latter technique are: Mg, O, and their compounds with B. From the commercial point of view the gaseous reaction is more expensive because very pure crystalline boron (99.7%) costs 16 $/g (Sigma-Aldrich 2011), whereas the magnesiothermic one produces amorphous B with a purity of 95-97% at a cost of about 2.655 $/g (Sigma-Aldrich 2011).

Doped boron is not yet commercially available, and thus it is necessary to prepare it on small quantity in the laboratory. Up to now, the doped boron market demand is not so high as to represent a tempting target for industries, because several unfavourable factors influence its price: the wide range of possible dopants and percentages, as well as the complexity of the doping techniques must to be reckoned with. Doping boron through "mechanical alloying" is unlikely; in fact, this element is practically only second to diamond on the Mohs's scale (hardness 9.3), so its grinding is difficult to implement. The chemical nature of boron and its doping elements, i.e. carbon (C), silicon carbide (SiC) and, more in general, elements or compounds with a high melting point, does not allow a direct dispersion because all these materials have poor metallic behaviour and high melting point (above 2076 °C).

The main goal of the present paper is to describe a novel technique to produce very fine and amorphous boron powders. The same technique could be useful to produce doped boron, to improve the $MgB_2$ upper critical field ($H_{c2}$) [11], [12], as well as a dispersion of the dopant in the $MgB_2$ matrix, to improve the pinning force by creating a regular lattice of defects (pinning centres) [13], [14]; in both cases an increase of the critical current flow in the conductor is obtained. We think that this process will be the most suitable technique to add organic molecules, like carbohydrate, to obtain a chemical substitution at the boron site and to disperse nano-particles, such as SiC or titanium $TiO_2$, in the $MgB_2$ matrix.

2. Experimental details: preparation of the nano-sized and amorphous boron for $mgb_2$ synthesis

The full procedure to obtain boron powders is articulated in three stages which will be summarized here and explained later (the whole process is sketched in Figure 1.):

1) Solubilization, cryogenic freezing and freeze-drying: this stage represents the innovative step to reduce the boron precursor size at a nanometric scale.

2) Reduction to row elemental boron, followed by acid leaching and the final heat treatment: this stage is the Moissan's process to refine and obtain the boron powder.

3) Synthesis of the $MgB_2$ powder for the *ex-situ* process.

*2.1. Solubilization*

The first step of the process is the solubilization of $B_2O_3$ in deionized water (Milli-Q®), purified and deionized to a high purity degree (resistivity 18.2 MΩ*cm). This dissolution, carried out at temperature very close to the water boiling point, produces boric acid and leads to a homogeneous dispersion of the molecules. Particular care needs to be taken in controlling the solution solubility limit in order to avoid crystallization phenomena.

*2.2. Cryogenic-freezing*

Once the precursor ($B_2O_3$ or doped- $B_2O_3$) dissolution is complete, the solution undergoes a *cryogenic-freezing* to transfer the liquid homogeneity in a solid lattice, thus obtaining very fine particles. The boiling solution is then sprayed into cryogenic liquid by a spray gun, which picks up the solution to be rapidly transferred into a liquid nitrogen ($LN_2$) vessel from the dissolution beaker. The direct contact between the sprayed solution and the cryogenic liquid creates a freezing of very fine particles which reflects the initial homogeneous composition of the liquid solution, and does not allow grain growth and/or segregation phenomena in the solid phase, should some doping elements be present.

*2.3. Freeze-drying*

The freeze-drying process (performed by a Coolsafe55, Scanvac) is applied to remove the water trapped during cryogenic-freezing; this maximizes the particle surface area and facilitates the solvent sublimation. This process is based on the sublimation principle, and is typically carried out at a T ≤ -55 °C, using a condensing chamber for the water vapour. The solid phase is placed in a suitable container linked to the condensation chamber and both volumes are kept under dynamic vacuum by a rotary pump. When the solvent used is water, the freeze-drying is carried out keeping the pressure below 611 Pa, corresponding to the water triple point. The direct solid-to-vapour phase transition ensures that the chemical homogeneity within the liquid solution will be kept in the final solid phase as well. Obviously, the finer the starting powders the finer and more porous the final powders. In fact, water sublimation from the solid solution produces many voids within the $B_2O_3$ aggregates [Figure 2(d)]. Fig. 2 shows $B_2O_3$ before [2(a) and 2(c)] and after [2(b) and 2(d)] the freezing-dry process. By comparing Fig. 2(c) with Fig. 2(d), at the same high magnification, we can clearly say that the freeze-dried $B_2O_3$ has a more amorphous and porous structure than the pristine $B_2O_3$.

*2.4. Reduction to elemental boron*

After the freeze-drying process had been completed, the product thus obtained ($B_2O_3$ or doped-$B_2O_3$ without water) was chemically reduced to elemental B. In order to remove residual moisture from the processed $B_2O_3$, the powders were thermally treated in a furnace under Ar flow. The reduction was carried out by mixing $B_2O_3$ with Mg (in ratio of 2.6:1 W/W) and reacting them in Ar flow in a furnace. Instead of using a slow heating rate, as in the Moissan's method, we reacted the powders at 1000 °C with a fast heating rate (around 50°C/min) in order to preserve the $B_2O_3$ characteristics impressed by freeze-drying and have a rapid reaction between precursors.

*2.5. Acid leaching and heat treatment*

As a consequence of B reduction from $B_2O_3$, the magnesium oxide by-product has to be removed: to this purpose we performed a series of acid leaching (using a 18% V/V solution of boiling hydrochloric acid), followed by water rinse until a neutral pH was reached.

In order to purify the B powder further, a final thermal treatment in Ar/5% $H_2$ flow for 12 h was applied and two different temperatures were tested: 700 °C and 900 °C. In the following the boron produced with the above described technique will be called $B_{NA}$ (Nano-Amorphous B), and the commercial boron will be $B_{HCS}$. Whole the process is described more in details in Ref. [15].

*F. $MgB_2$ synthesis and tapes preparation*

Monofilamentary tapes were fabricated following the *ex-situ* route of the conventional P.I.T. method [16]. $MgB_2$ powders, using both commercial boron (amorphous boron grade I purchased from H. C. Starck or $B_{HCS}$) and $B_{NA}$ heat treated at 700 °C, were synthesized at 900 °C for 1 hour in Ar flow and packed inside Ni tube of 12.7 mm external and 6 mm internal diameter. The tapes were groove rolled and drawn down to a 2 mm diameter wire, then cold rolled in several steps to a tape of about 0.38 mm in thickness and 4 mm in width, with a filling factor of 25-30%. The superconducting transverse cross section of the conductor was about 0.4 mm². The conductors were then subjected to a final sintering treatment at 920°C for 20 min in flowing argon in order to recover the stress collected during the cold working and sinter $MgB_2$ grains.

*2.6. Structural characterization of boron*

About 50 g of $B_{NA}$ were prepared for this study. The grain size and morphology was evaluated using a scanning electron microscope (SEM), by acquiring several surface area images representative of the sample at different magnification. We analyzed the presence of impurity and the boron phase before and after the final heat treatment (step G) by x-ray. The same analysis technique was applied to establish the effect of the final thermal process. The θ-2θ scans of different B phases are reported in Fig. 3 and 4. Transmission electron microscopy (TEM) analysis was carried out on the powders using a JEOL JEM-2010 operated at 200 kV; the analytical electron microscope (AEM) investigation was performed using an energy-dispersive x-ray spectrometer system (EDS: OXFORD PENTAFET). The specimens for TEM examination were prepared by gently milling the reacted powders that were subsequently ultrasonically dispersed in n-pentanol and then deposited onto a Cu support grid coated with a holey carbon film.

*2.7. Magnetic and electrical characterization of tapes*

$MgB_2$ phase was checked in the *ex-situ* powders by Xrd technique using a PHILIPS diffractometer (Bragg-Brentano geometry, Ni filtered Cu Kα radiation). Short pieces of tapes, about 6 mm in length, were employed for magnetization vs. temperature measurements with a commercial 5.5 MPMS Quantum Design Squid magnetometer. Transport critical current measurements were performed on 10 cm long tapes at the Grenoble High Magnetic Field Laboratory at 4.2 K in magnetic field up to 9 T, applied perpendicular to the tape surface, while the current was applied perpendicular to the field.

3. Results and discussion

*3.1. Structural characterization of boron*

Fig. 2 shows SEM pictures of boron oxide before [left panel (a), (c)] and after [right panel (b), (d)] the freeze-drying process. The lyophilized $B_2O_3$ powder needs a higher magnification to show its amorphous structure. On the contrary, untreated (not lyophilized) $B_2O_3$ shows the same structure at lower and higher magnification. Before the lyophilization process, the boron oxide looks like a dense overlapping of planar structures with hexagonal shape. After it, we can always observe a hexagonal structure, where smaller aggregates show many voids within the hexagonal plane. The voids are due to water sublimation from the solid system $H_2O$-$B_2O_3$. The porous structure is suitable to give the higher reaction surface to magnesium during the $B_2O_3$ reduction. In fact, in this form the $B_2O_3$ is thoroughly intimately mixed with Mg powders. The freeze-dried $B_2O_3$ was then reacted with Mg and raw boron was purified from MgO, as described before. The quality of the B powder obtained was investigated by Xrd analysis in order to detect secondary phases, such as MgO, $B_2O_3$ and other impurities. The Xrd technique was also employed to understand the effect of temperature on the final B phase. Fig. 3 shows Xrd patterns for the $B_{NA}$ before any heat treatment, and boron heat treated at 700 °C and 900 °C, respectively. Here and in Fig. 4, the theoretical diffraction pattern of boron is shown as a reference (Pearson's Crystal Data n° 1700899) [17]. From the XRD patterns it is evident that, by tuning temperature in the range of 700 and 900 °C, it is possible to change the boron phase from amorphous to crystalline one. Before the treatment, boron seems pure and amorphous ,whereas, after the heat treatment at 900 °C some $B_2O_3$ phase emerges. This could be due to $B_2O_3$ impurities not completely removed by leaching process and/or to the nano-sized amorphous boron oxidization when exposed to air for characterization. As to the latter, we can only exclude oxygen contamination during heat treatment, since the furnace used is directly linked to a glove box and the thermal process is in Ar/ 5% $H_2$ flow, as already described in a previous work [7]. The first source, on the other hand, might be due to an incomplete reduction of $B_2O_3$. In fact, during the reduction from boron oxide to boron, the liquid Mg reacting with $B_2O_3$ can form a MgO crust, and the un-reacted $B_2O_3$ core turns out to be shielded from further reduction. If this is the case, acid-leaching during the washing process will be vain. In order to avoid this problem, it could be useful to improve the precursor mixing and/or sieving the $B_2O_3$ after the lyophilisation process. This oxidizing issue needs further study to understand the different sources of $B_2O_3$. In this regard it could be interesting to investigate *in-situ* the $B_2O_3$ formation during the various steps of the process by the high-energy Xrd technique used in a latter work [18]. Fig. 4 compares Xrd patterns of $B_{NA}$ at 700°C and 900°C with the commercial one, nominally amorphous. It is clear from this comparison that $B_{HCS}$ is more crystalline than the $B_O$ boron. In order to confirm the Xrd results, we performed a TEM analysis on the $B_{NA}$ and $B_{HCS}$ powders, showed in Fig. 5.

We can observe a really amorphous structure corresponding to $B_{NA}$ treated at 700 °C Fig. 5(a). Fig. 5(b), relative to $B_{HCS}$, shows lattice fringes typical of a crystalline boron structure; the inset shows the corresponding electron diffraction pattern with the characteristic hexagonal symmetry. Both the B powders ($B_{NA}$ at 700 °C and $B_{HCS}$) were observed by scanning electron microscope. Fig. 6 shows a representative SEM image, chosen among the nine taken, of $B_{NA}$ and $B_{HCS}$. From these pictures, it is evident that $B_{NA}$ powders are more amorphous and more homogeneously fine then the commercial one. Both the images are at the same magnification (10000X). From morphological and grain size point of view the SEM image relative to the boron powder treated at 900 °C looks very similar to that of the 700 °C heat treated powders. Thus, there is no need to show them. All the nine SEM images were used for statistical count of grain size distribution using a Matlab routine. The Lorentzian fits of the raw data obtained are shown in Fig. 7. As can be seen, our B is much finer than the commercial one (55 nm Vs 158 nm), and has a better homogeneity (310 nm Vs 865 nm).

*3.2. MgB$_2$ tapes characterization*

Fig. 8 shows the comparison between Xrd patterns of $MgB_2$ powders synthesized using $B_{HCS}$ and $B_{NA}$ $MgB_2$ obtained from $B_{NA}$ precursor has a little amount of elemental Mg and a higher MgO content with respect to $MgB_2$ obtained from commercial precursors. The un-reacted Mg amount might be due to the micrometric nature of the elemental Mg, which could need more time to react with the nanometric and amorphous B than the commercial one. This is another issue we must to investigate more in detail for future developments. The Xrd pattern analysis corresponding to the $MgB_2$ obtained from $B_{NA}$ precursor does not show any shift in the $MgB_2$ peaks position. Fig. 9 presents the normalized magnetic moment as a function of temperature obtained at 1 mT for the two $MgB_2$ tapes made with the two kind of boron. In the following we refer to them simply as "lyophilized tape" and "commercial tape". The onset of the critical temperature of the lyophilized tape is 37 K, whereas the commercial one has 38 K. In order to explain this difference we have to consider that, generally, the superconducting powders inside a conductor have a lower $T_C$ than the same powders in a bulk shape: this is due to the lattice stress experienced by powders within the tape during the cold working, which can be recovered by applying a proper final sintering heat treatment [19], [20]. Starting from the finest B powders, $MgB_2$ powders obtained from $B_{NA}$, nano-sized and amorphous, are finer than $MgB_2$ synthesized from commercial boron. As a consequence of their nano-sized nature, these powders show a different behaviour during cold deformation. They probably collect more stress compared to the standard $MgB_2$ powders, and therefore the corresponding $T_C$ is more affected. Furthermore, $MgB_2$ powders obtained from our boron, being very fine and homogeneous in size, are not easily packaged within the Ni sheath, and often the Ni tube gets broken during cold working. This does not happen to the standard powder produced using $B_{HCS}$ precursor. Just to give an idea on how much different the two $MgB_2$ powders are, we report the powders density inside the Ni tube after their packaging. We calculated the density by measuring the internal volume of the Ni tube and weighting the powder mass. Using a hydraulic press to package powder within the Ni sheath and a packaging pressure of $2.07*10^6$ Pa for the "commercial tape", we obtain a density value of 1.15 g/cm$^3$. Instead, when we use MgB$_2$ powders synthesized from B$_{NA}$ the resulting density is not greater than 1.05 g/cm$^3$, although a higher pressure of $5.52*10^6$ Pa is used. We think that many parameters still need to be developed to optimize these powders as a function of the P.I.T. process, of the sheath material choice and of the final sintering process. The transverse cross section of two tapes is given in Fig. 10. At the Ni/MgB$_2$ interface the tape filled with the standard MgB$_2$ powders shows a better deformed shape than the "lyophilized tape". Fig. 11 shows the transport critical current density ($J_{C\,t}$) as a function of magnetic field at 4.2 K for the two tapes. The "lyophilized tape" presents a better behaviour for high fields (> 2.5 T) while the "commercial tape" shows a better behaviour in the low-field region, even though data below 3.5 T were not taken because of overheating due to absence of thermal stabilization. This $J_{C\,t}$ increase at high field can be due to upper critical field and/or to pinning enhancement but, at the moment, we don't have enough data points to extract this information, and it is not the aim of this paper.

4. Conclusions

In the present paper a novel method to produce very fine and amorphous boron by magnesiothermic reaction, starting from a water solution containing soluble B$_2$O$_3$ only, has been reported. At the end of the above described process a very fine and amorphous B has been obtained; which could be, in our opinion, a suitable precursor to minimize the synthesis parameters, such as time and temperature involved in the chemical reaction to synthesize MgB$_2$. This paper represents the starting step of a more complicated process, which could be useful to produce other interesting product, i.e. doped boron, by adding several soluble and/or insoluble doping agents at the beginning of the process, first step of stage one in to the Fig. 1. More measurements in order to clarify this and other points should be performed. Furthermore, the mechanical deformation process must be improved and adapted on the our powder, so that all the potential of MgB$_2$ can be emphasized. In fact, the tapes reported in the present paper have been both obtained by a standard deformation process followed by a standard final sintering treatment [19], which is widely optimized up to now only for the commercial powders.

Acknowledgments
The authors wish to acknowledge the financial support of the Italian Miur project (PRIN 2008), Shrikant Kawale for his scientific support and Xavier Chaud from GHMFL (Grenoble) for technical support.

**Figure captions**

Fig. 1. Sketch of the novel procedure to produce nanosized and amorphous boron: Step 1 is the novel procedure applied before magnesiothermal process, step 2 represents the Moissan's method to produce boron, and step 3 is the $MgB_2$ synthesis.

Fig. 2. SEM images of boron oxide before [left panel: (a), (c)] and after [right panel: (b), (d)] the freeze-drying process at lower and higher magnification.

Fig. 3. XRD patterns relative to $B_{NA}$ before the final heat treatment (2) and after the thermal treatment at 700 °C (3) and 900 °C (4). Curve (1) represents the theoretical XRD pattern of crystalline B (Pearson's Crystal Data card n° 1700899).

Fig. 4. Comparison between different boron powders: $B_{NA}$ treated at 700 °C (2) and 900 °C (3), $B_{HCS}$ (4). Curve (1) represents the theoretical XRD pattern of crystalline B (Pearson's Crystal Data card n° 1700899).

Fig. 5. HRTEM images of $B_{NA}$ (a) and $B_{HCS}$ (b) boron; insets show the corresponding selected area electron diffraction patterns.

Fig. 6. SEM pictures at the same magnification of $B_{NA}$ treated at 700 °C (a) and $B_{HCS}$ (b).

Fig. 7. Grain size distribution of $B_{HCS}$ powders (triangle) and $B_{NA}$ (circle), measured by image treatment software from SEM images and fitted with a Lorentzian curve.

Fig. 8. XRD patterns of $MgB_2$ powders as synthesized using $B_{NA}$ and $B_{HCS}$.

Fig. 9. Normalized magnetic moment obtained at 1mT as a function of temperature for $MgB_2$ "commercial tape" (triangle) and "lyophilized tape" (circle). The inset shows a magnification close to $T_C$.

Fig. 10. Figure 10. Transverse cross section of a $MgB_2$: (a) "lyophilized tapes" and (b) "commercial tapes"

Fig. 11. Transport critical current density of $MgB_2$ tapes obtained from $B_{HCS}$ (triangle) and $B_{NA}$ (circle) boron. The measurements were performed at 4.2 K with the tape perpendicular to the magnetic field.

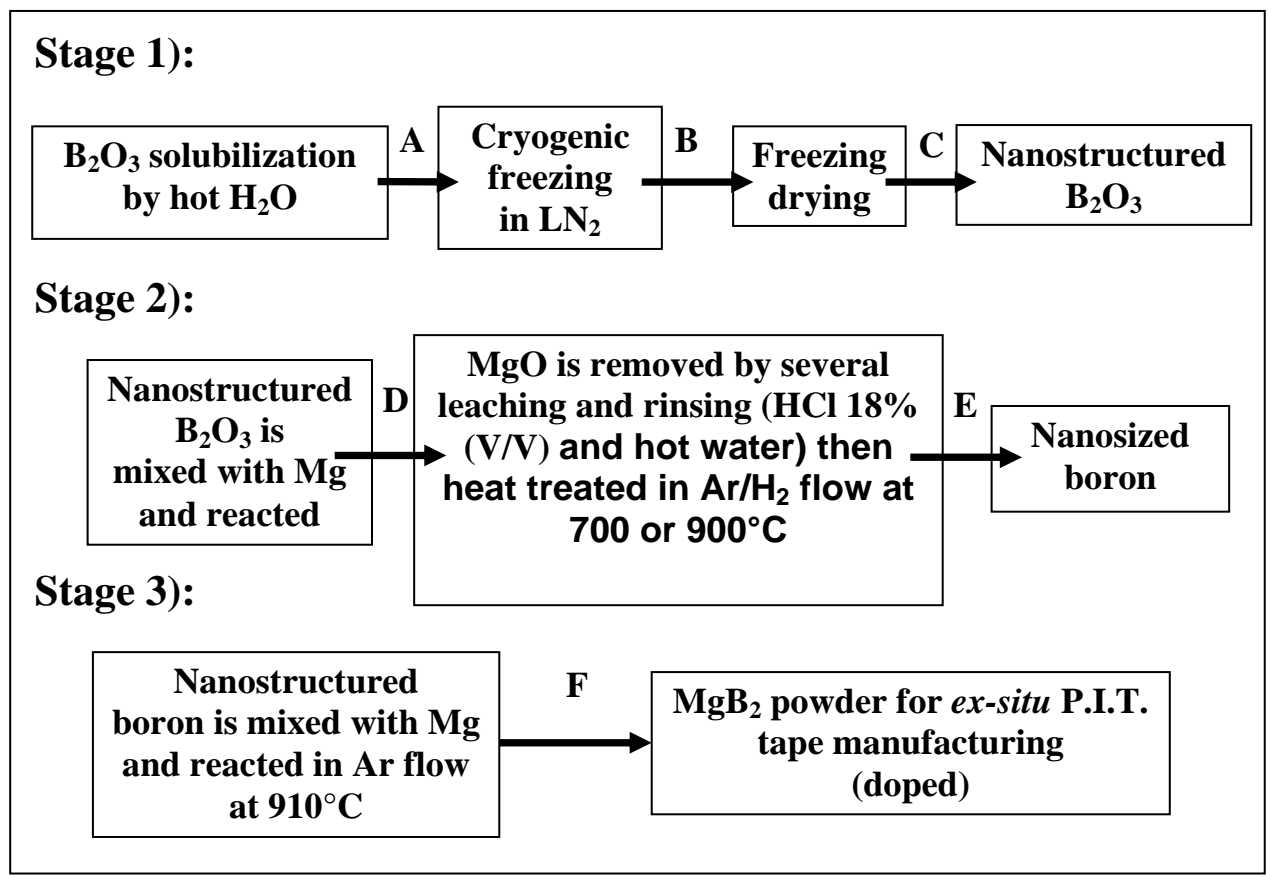

Fig. 1. Sketch of the novel procedure to produce nanosized and amorphous boron: Step 1 is the novel procedure applied before magnesiothermal process, step 2 represents the Moissan's method to produce boron, and step 3 is the $MgB_2$ synthesis.

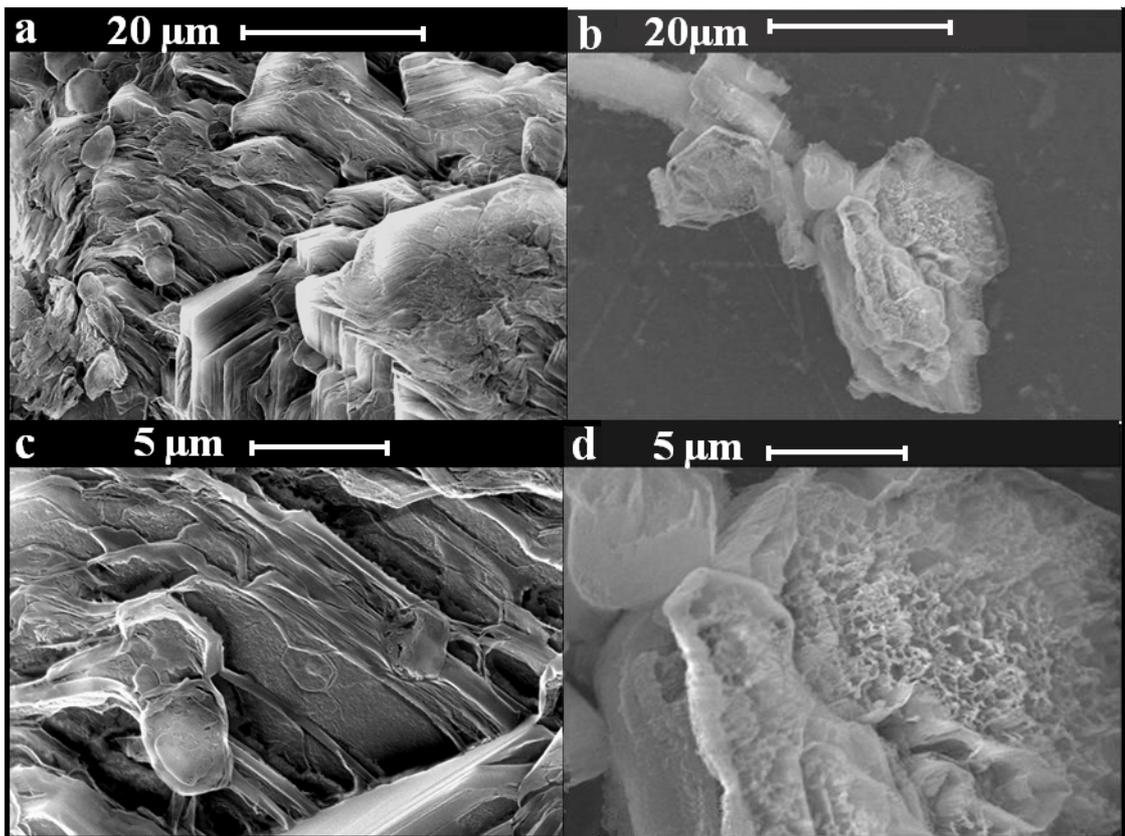

Fig. 2. SEM images of boron oxide before [left panel: (a), (c)] and after [right panel: (b), (d)] the freeze-drying process at lower and higher magnification.

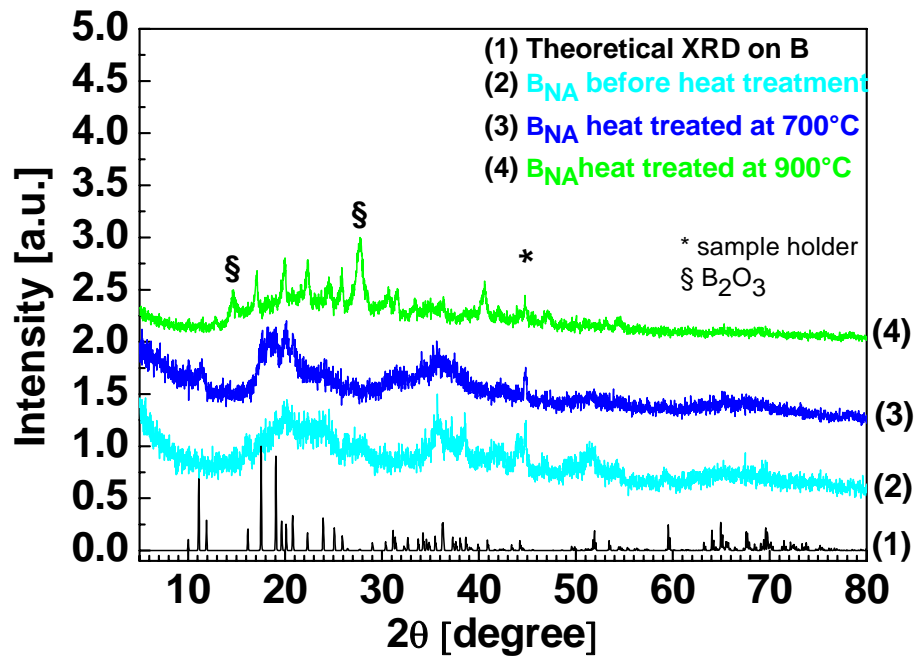

Fig. 3. XRD patterns relative to the $B_{NA}$ before the final heat treatment (2) and after the thermal treatment at 700 °C (3) and 900 °C (4). Curve (1) represents the theoretical XRD pattern of crystalline B (Pearson's Crystal Data card n° 1700899).

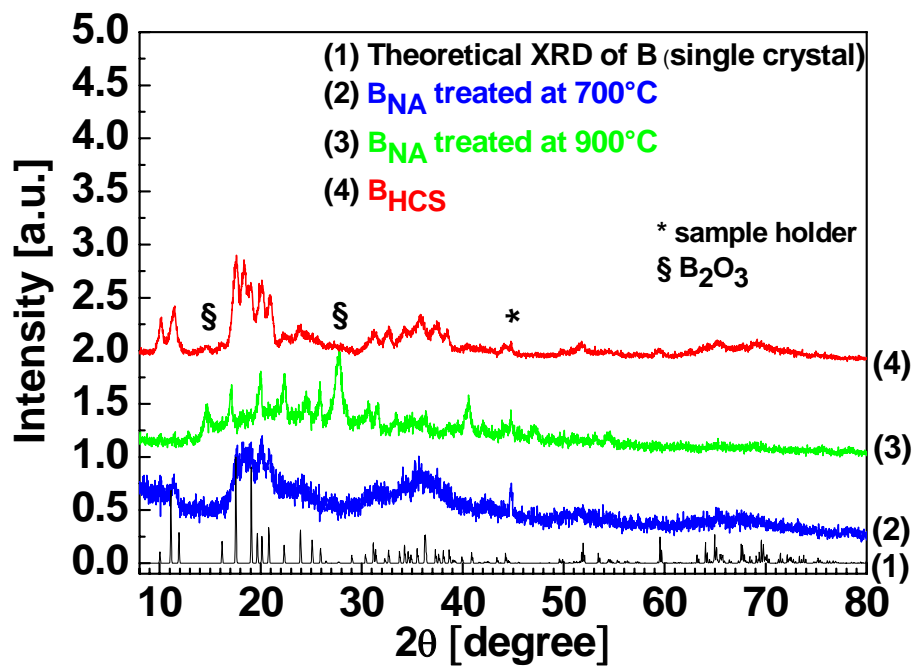

Fig. 4. Comparison between different boron powders: $B_{NA}$ treated at 700 °C (2) and 900 °C (3), $B_{HCS}$ (4). Curve (1) represents the theoretical XRD pattern of crystalline B (Pearson's Crystal Data card n° 1700899).

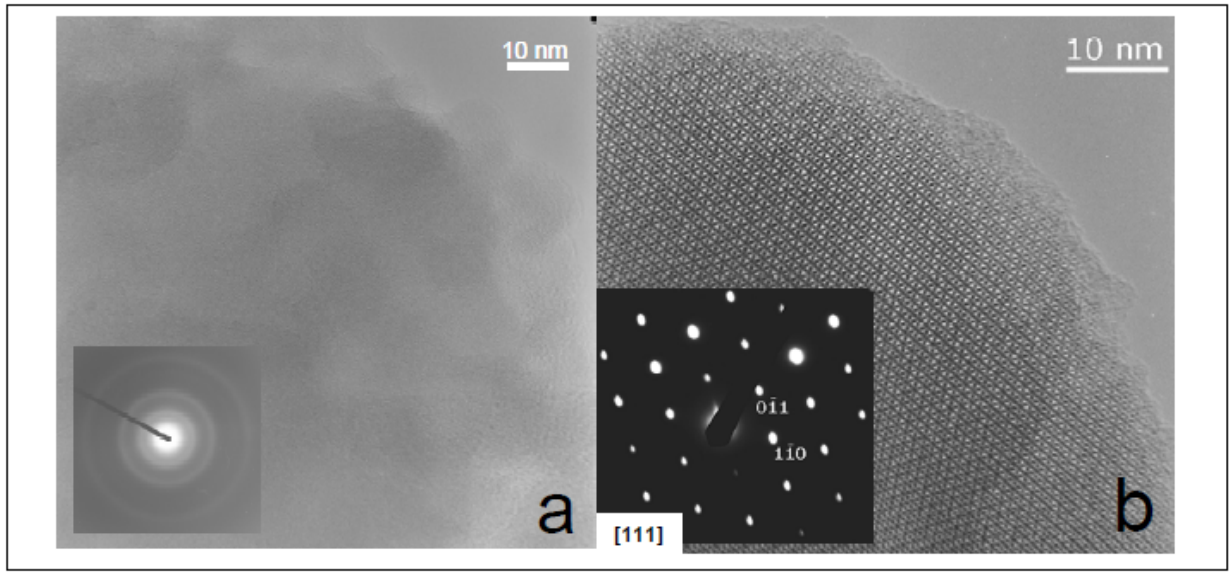

Fig. 5. HRTEM images of $B_{NA}$ (a) and $B_{HCS}$ (b) boron; insets show the corresponding selected area electron diffraction patterns.

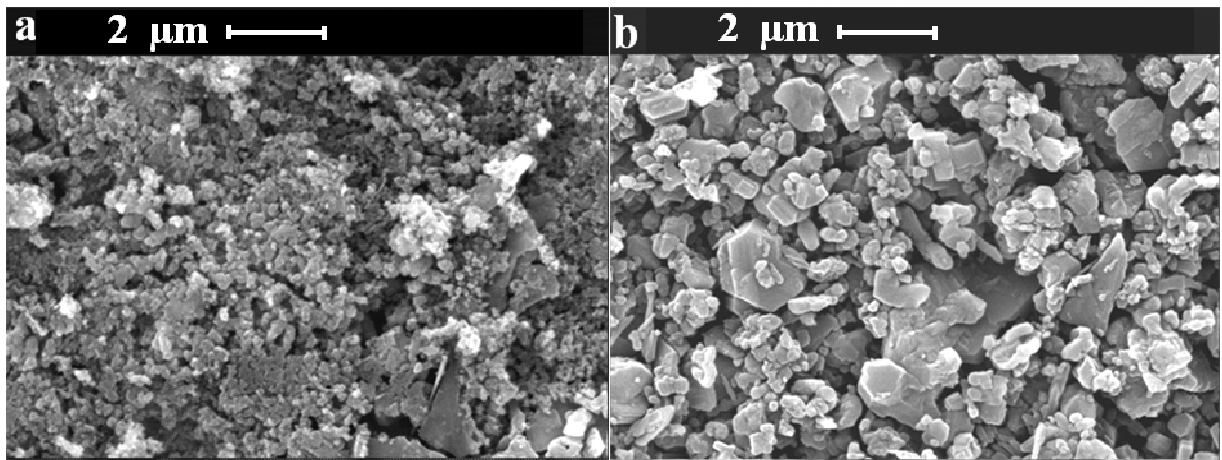

Fig. 6. SEM images of (a) $B_{NA}$ treated at 700°C and (b) $B_{HCS}$ at the same magnification.

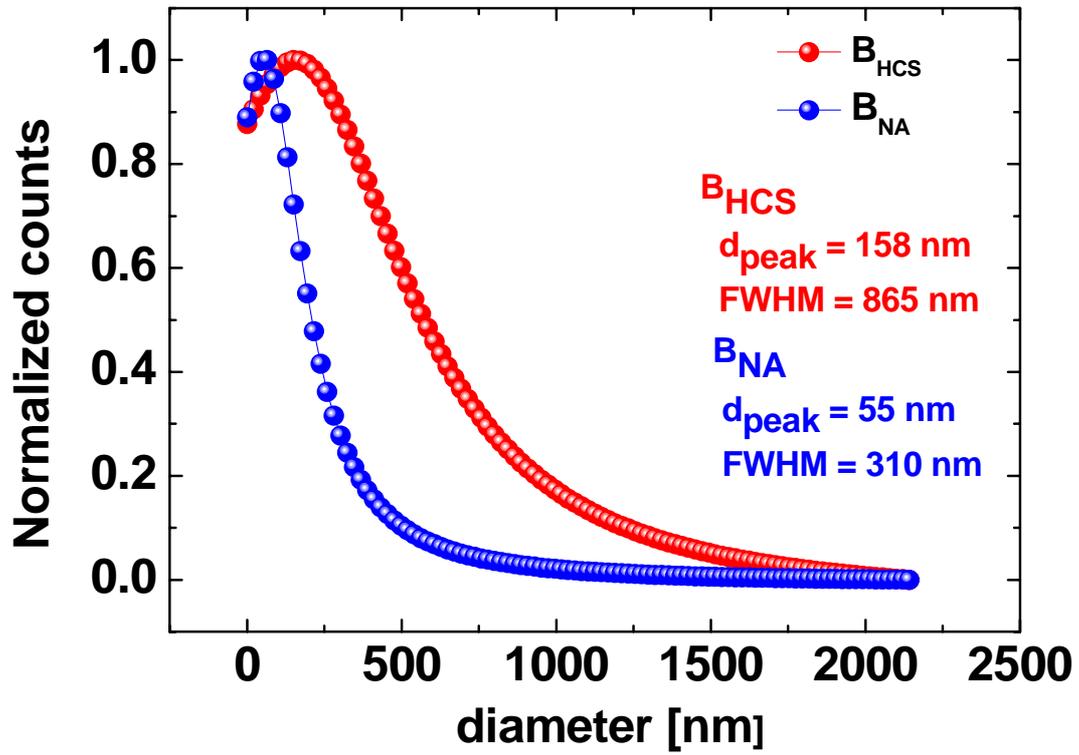

Fig. 7. Grain size distribution of $B_{HCS}$ powders (triangle) and $B_{NA}$ boron (circle), measured by an image treatment software on SEM images and fitted with a Lorentzian curve.

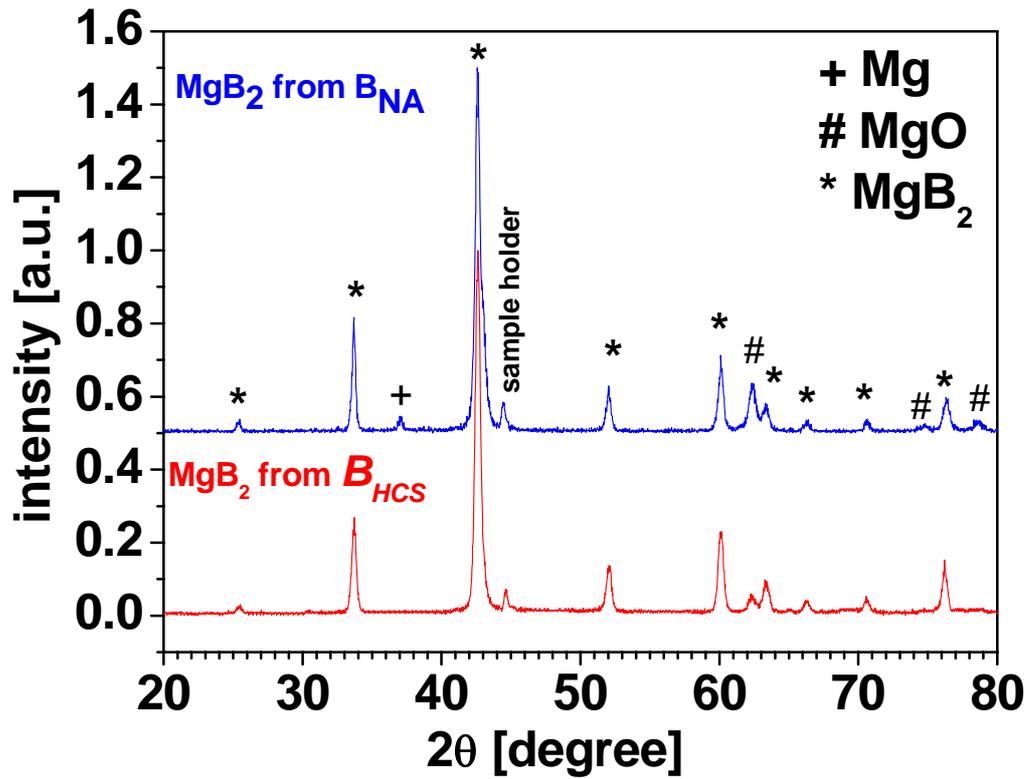

Fig. 8. XRD patterns of MgB$_2$ powders as synthesized using B$_{NA}$ and B$_{HCS}$.

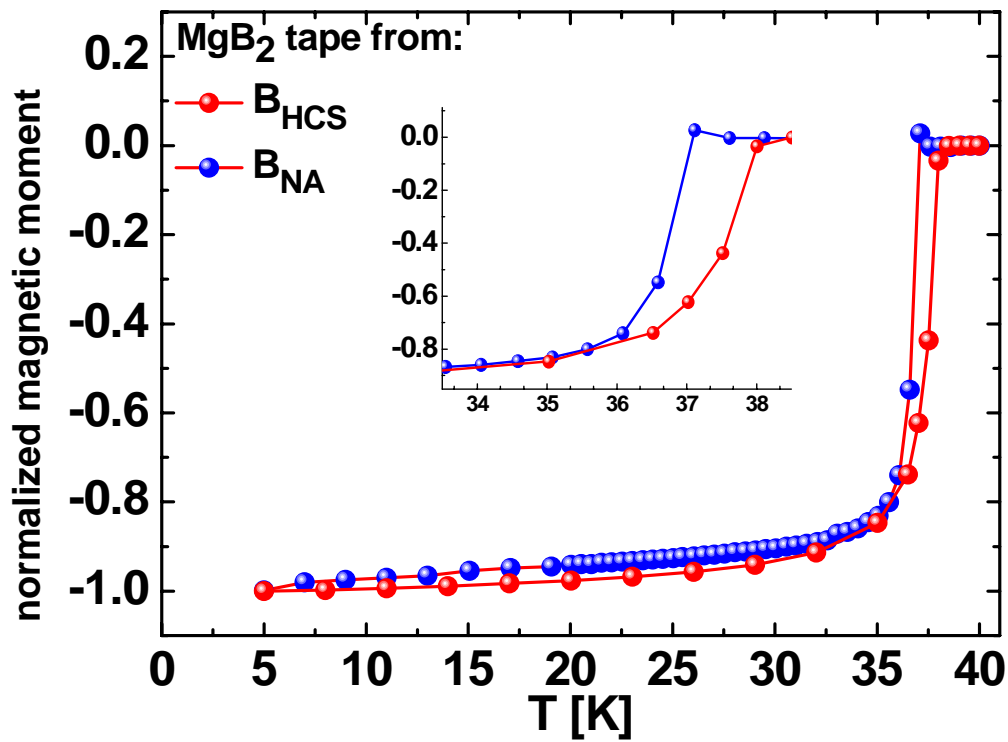

Fig. 9. Normalized magnetic moment obtained at 1 mT as a function of temperature for MgB$_2$ tapes made with B$_{HCS}$ (triangle) and B$_{NA}$ (circle). The inset shows a magnification close to $T_C$.

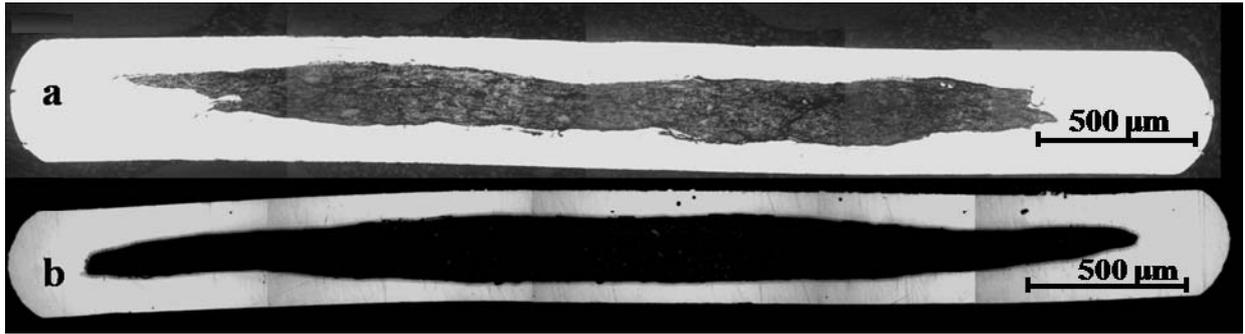

Fig. 10. Transverse cross section of a: (a) MgB$_2$ tape filled with powder synthesized from B$_{NA}$ and (b) MgB$_2$ tapes filled with powder synthesized from B$_{HCS}$.

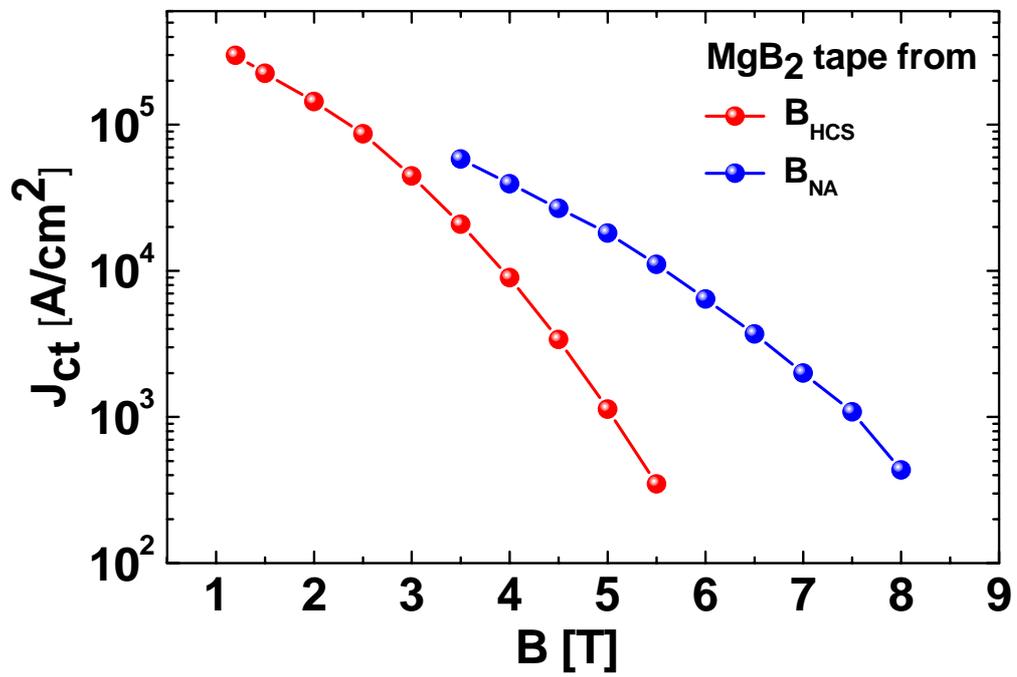

Fig. 11. Transport critical current density of MgB$_2$ tapes obtained from B$_{HCS}$ (triangle) and B$_{NA}$ (circle). The measurements were performed at 4.2 K with the tape perpendicular to the magnetic field.